\documentclass[fleqn,twoside]{article}
\usepackage[headings]{espcrc2}

\readRCS
$Id: espcrc2.tex,v 1.2 2004/02/24 11:22:11 spepping Exp $
\usepackage{graphicx}
\usepackage[figuresright]{rotating}

\newcommand{\AmS}{{\protect\the\textfont2
  A\kern-.1667em\lower.5ex\hbox{M}\kern-.125emS}}

\title{Exclusive Processes at Colliders}

\author{A. Szczurek \address{Institute of Nuclear Physics, PL-31-342 Cracow, Poland, \\ University of Rzesz\'ow, PL-35-959 Rzesz\'ow, Poland}}%
       
\runtitle{2-column format camera-ready paper in \LaTeX}
\runauthor{A. Szczurek}

\begin{document}

\begin{abstract}
A few examples of exclusive processes at high energy are discussed.
Several mechanisms are presented. The differential distributions are
shown. The possibilities to measure the processes are discussed.
\vspace{1pc}
\end{abstract}

\maketitle

\section{Introduction}

Up to now the investigations at high energies
were concentrated on inclusive processes, i.e.
processes (measurements) when only one object of many produced
simultaneously is recorded. Naively, these processes seem to be
difficult but can be well described within the standard parton
model. The exclusive processes seems naively easier, but in reality
require detailed knowledge of the QCD dynamics, which is 
not required in the inclusive case.

The exclusive production was studied in detail
mostly close to the kinematical thresholds.
The Tevatron opened a possibility to study the central 
(semi)exclusive production of mesons (elementary objects)
at high energies. 
A similar program will be carried out in the future 
at the LHC.
Here I review a few examples of exclusive
production I have studied recently with my 
collaborators
(for details see Refs. 
\cite{SPT07,SS07,PST08,PST09,RSS08,SL08,LS09,KSS09}).

\begin{figure}
\begin{center}
  \includegraphics[height=.25\textheight]{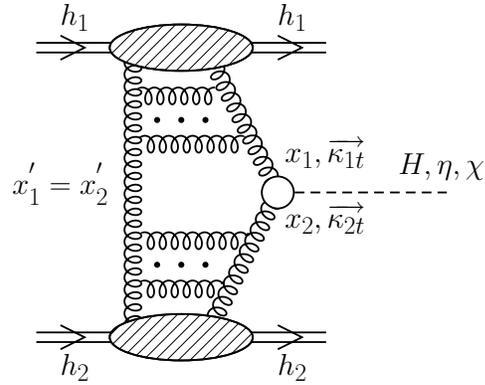}
\vspace{-0.7cm}
\end{center}
  \caption{A sketch of the bare QCD mechanism of
exclusive heavy scalar meson production.
\label{fig:scalar_diagram}
}
\end{figure}

\begin{figure}
\begin{center}
  \includegraphics[height=.2\textheight]{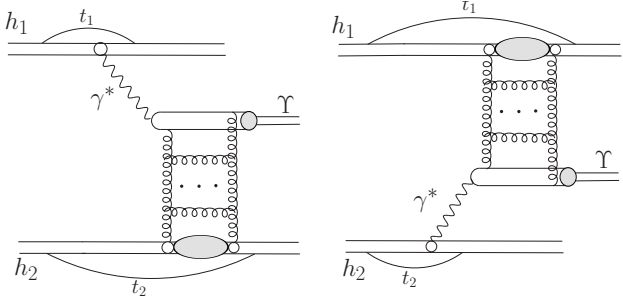}
\vspace{-0.7cm}
\end{center}
  \caption{Two basic QED $\otimes$ QCD mechanisms of
exclusive heavy vector meson production.
\label{fig:vector_diagram}
}
\end{figure}

For exclusive production of a single object (Higgs or meson), 
the mechanism of the reaction depends
on the quantum numbers of the object and/or its internal
structure.
For heavy scalar mesons (scalar quarkonia, scalar glueballs) 
the mechanism of the production, shown in 
Fig.\ref{fig:scalar_diagram}, 
is the same as for the diffractive Higgs boson 
production extensively discussed in recent years
\cite{KMR}.
The dominant mechanism for the exclusive heavy 
vector meson production is quite different.
Here the dominant processes are shown in 
Fig.\ref{fig:vector_diagram}. These processes were discussed by
A. Cisek during this conference \cite{cisek}.

\begin{figure}
\begin{center}
  \includegraphics[height=.25\textheight]{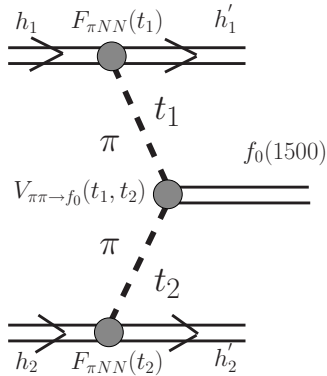}
\vspace{-0.7cm}
\end{center}
  \caption{A sketch of the bare QCD mechanism of
exclusive heavy scalar $f_0(1500)$ meson production.
\label{fig:pion_pion_diagram}
}
\end{figure}

When going to lower energies the mechanism of the meson production 
becoming more complicated and the number of mechanisms increases. 
For example, in Fig.\ref{fig:pion_pion_diagram} 
I show the mechanism of the glueball candidate $f_0(1500)$ production 
which plays dominant role at low energies \cite{SL08}.


\section{Different examples}

\subsection{Exclusive production of $\chi_c$ mesons}

The amplitude of the exclusive double diffractive color singlet
production $pp\to pp\chi_{cJ}$ can be written as \cite{KMR}:
\begin{eqnarray}
{\cal
M}^{g^*g^*}=\frac{s}{2}\cdot\pi^2\frac12\frac{\delta_{c_1c_2}}{N_c^2-1}\,
\Im\int
d^2 q_{0,t}V^{c_1c_2}_J \nonumber \\
\frac{f^{off}_{g,1}(x_1,x_1',q_{0,t}^2,
q_{1,t}^2,t_1)f^{off}_{g,2}(x_2,x_2',q_{0,t}^2,q_{2,t}^2,t_2)}
{q_{0,t}^2\,q_{1,t}^2\, q_{2,t}^2} \; .
\label{ampl}
\end{eqnarray}
The amplitude is averaged over the color indices and
over the two transverse polarizations of the incoming gluons \cite{KMR}.

In calculating the vertex $V^{c_1c_2}_J$ we have included
off-shellness of gluons \cite{PST08}. The unintegrated
gluon distributions were taken from the literature.
We have shown in Ref.\cite{PST08} that 
for relatively light $\chi_c(0)$, unlike for
the Higgs boson \cite{KMR}, the dominant 
contributions come from the nonperturbative regions 
of rather small gluon transverse momenta.

\begin{figure}[!h]    
\includegraphics[width=0.4\textwidth]{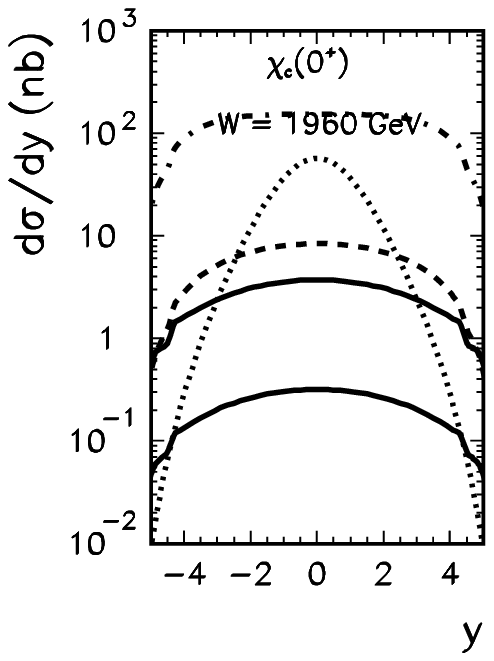}
\includegraphics[width=0.4\textwidth]{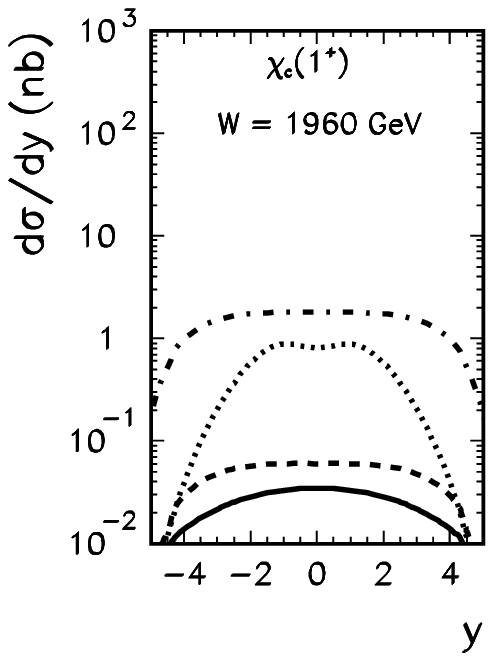}
\vspace{-0.7cm}
   \caption{\label{fig:dsig_dy}
   \small Distribution in rapidity of $\chi_c(0^+)$ meson
(upper panel) and $\chi_c(1^+)$ meson (lower panel) for
different UGDFs.
}
\end{figure}

In Ref.\cite{PST08} we have made a detailed presentation
of differential distributions. Here I show only distribution
in rapidity of $\chi_c(0^+)$ (see Fig.\ref{fig:dsig_dy}).
Although all UGDFs give a similar quality description of
the low-$x$ HERA data for the $F_2$ structure function,
they give quite different rapidity distributions of
$\chi_c(0^+)$.
The UGDFs which take into account saturation effects (GBW, KL) give much
lower cross section than the BFKL UGDF (dash-dotted line).
Therefore the process considered here would help, at least 
in principle, to constrain rather poorly known UGDFs.

There is interesting theoretical aspect of the double diffractive
production of the $\chi_c(1^{+})$ meson. The coupling 
$g g \chi_c(1^{++})$ vanishes for on-shell
gluons (so-called Landau-Yang theorem). According to the original
Landau-Yang theorem \cite{LY_theorem} the symmetries under space
rotation and inversion forbid the decay of the spin-1 particle into
two (on-shell) spin-1 particles (two photons, two gluons). The same
is true for the fusion of two on-shell gluons. The symmetry
arguments cannot be strictly applied for off-shell gluons. 

In Ref.\cite{PST09} we have shown explicitly that the Landau-Yang 
theorem is violated by virtual effects in diffractive production 
of $\chi_c(1^+)$ leading to important consequences. 
In our approach the off-shell effects are treated explicitly. 
The measurement of the cross section can be therefore a good test 
of the off-shell effects and consequently UGDFs used in 
the calculation.

In Fig.~\ref{fig:dsig_dy} I show distributions
in rapidity $y$ of $\chi_c(1^+)$ for different UGDFs from 
the literature.
The results for different UGDFs differ significantly. The biggest
cross section is obtained with BFKL UGDF and the smallest cross
section with GBW UGDF. The big spread of the results is due to quite
different distributions of UGDFs in gluon transverse momenta.


The cross section for the axial-vector $\chi_c(1^+)$ production 
is much smaller (about two orders of magnitude)
than the cross section for the scalar $\chi_c(0^+)$
production. This smallness can be understood in the context of
Landau-Yang theorem, which "causes" vanishing of
the cross section for on-shell gluons.

The $\chi_c$ mesons are usually measured through the observation of
the $\gamma + J/\Psi$ decay channel. The axial-vector $\chi_c(1^+)$ meson 
has a large branching fraction for radiative decay $\chi_c(1^+) \to \gamma
+ J/\psi$ (BR = 0.36 \cite{PDG}). This is significantly bigger than 
for the scalar $\chi_c(0^+)$ where it is only about 1 \% \cite{PDG}.
Therefore the discussed off-shell efects are very important to
understand the situation in the $\gamma + J/\Psi$ channel observed
experimentally. Full analysis requires inclusion of the $\chi_c(2^+)$ 
meson where the branching ratio is also relatively high.

\subsection{Exclusive production of the gluball candidate
$f_0(1500)$}

In Ref.\cite{SL08} we have discussed exclusive
production of scalar $f_0(1500)$ in the following 
reactions:
\begin{eqnarray}
&&p + p \to p + f_0(1500) + p \; , \\
&&p + \bar p \to p + f_0(1500) + \bar p \; , \\
&&p + \bar p \to n + f_0(1500) + \bar n \; . 
\label{f0(1500)_reactions}
\end{eqnarray}
While the first process could be measured at the J-PARC 
complex being completed recently or by the COMPASS collaboration, 
the latter two reactions could be measured by the PANDA Collaboration
at the new complex FAIR planned in GSI Darmstadt.
The combination of these processes could provide more information
on the mechanism of $f_0(1500)$ production and hopefully some information 
on its nature.



\begin{figure}    %
\includegraphics[width=5cm]{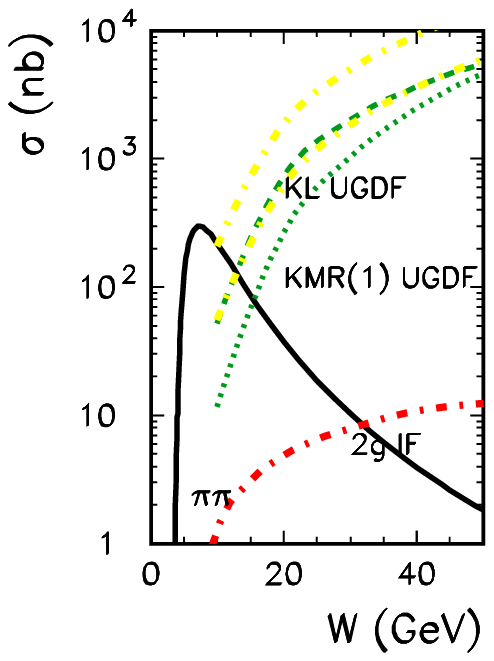}
\includegraphics[width=5cm]{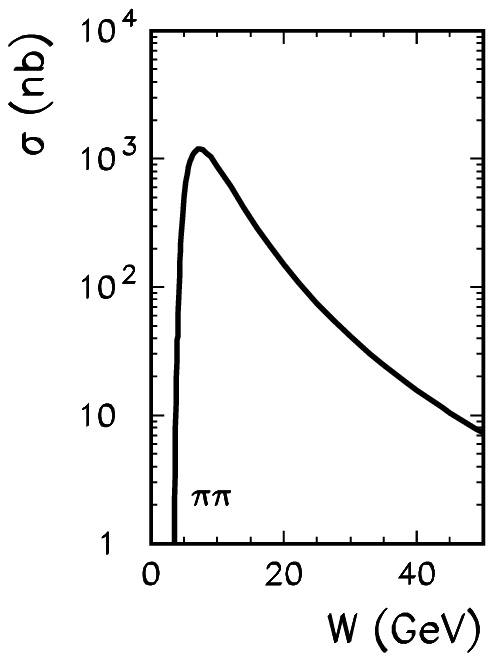}
\vspace{-0.7cm}
\caption{\label{fig:sigma_W}
The integrated cross section as a function
of the center of mass energy for $p \bar p \to p \bar p f_0(1500)$ 
(upper panel)
and $p \bar p \to n \bar n f_0(1500)$ (lower panel) reactions.
The thick solid lines are for pion-pion MEC contribution 
($\Lambda$ = 0.8, 1.2 GeV), the dashed line is 
for QCD diffractive contribution obtained with the Kharzeev-Levin UGDF, 
the dotted line for the KMR approach
and the thin solid lines (blue on-line) are for "mixed" UGDF
(KL $\otimes$ Gaussian) with $\sigma_0$ = 0.5, 1 GeV. 
The dash-dotted line represents the two-gluon impact factor 
result \cite{SL08}.
}
\end{figure}


In Ref.\cite{SL08} we have proposed a new mechanism 
(see Fig.\ref{fig:pion_pion_diagram}) which becomes 
dominant at lower energies.
In Fig.\ref{fig:sigma_W} we show the integrated cross section
for the exclusive $f_0(1500)$ elastic production
$p \bar p \to p f_0(1500) \bar p$ 
and for double-charge-exchange reaction
$p \bar p \to n f_0(1500) \bar n$.
The thick solid line represents the pion-pion component calculated with 
monopole vertex form factors with 
$\Lambda$ = 0.8 GeV (lower) and $\Lambda$ = 1.2 GeV (upper).
The difference between the lower and upper curves represents uncertainties
on the pion-pion component.
The pion-pion contribution grows quickly from the threshold, takes
maximum at $W \approx$ 6-7 GeV and then slowly drops with increasing
energy. The gluonic contribution calculated with unintegrated
gluon distributions drops with decreasing energy 
towards the kinematical threshold and seems to be about order of 
magnitude smaller than the pion-pion component at W = 10 GeV.
We show the result with Kharzeev-Levin UGDF (dashed line) which 
includes gluon saturation effects relevant for small-x, 
Khoze-Martin-Ryskin UGDF (dotted line) used for the exclusive 
production of the Higgs boson and the result with the 
"mixed prescription" (KL $\otimes$ Gaussian) \cite{SL08} 
for different values of the $\sigma_0$ parameter: 
0.5 GeV (upper thin solid line),
1.0 GeV (lower thin solid line). 
In the latter case results strongly depend on 
the value of the smearing parameter.

\subsection{Exclusive production of the $\pi^+ \pi^-$ 
pairs}

Up to now I have discussed only exclusive production of a 
single meson. Also the channels with meson pairs
seem interesting. In particular, the channel with
two charged pions which seems feasible experimentally.


\begin{figure}
\begin{center}
  \includegraphics[width=4.5cm]{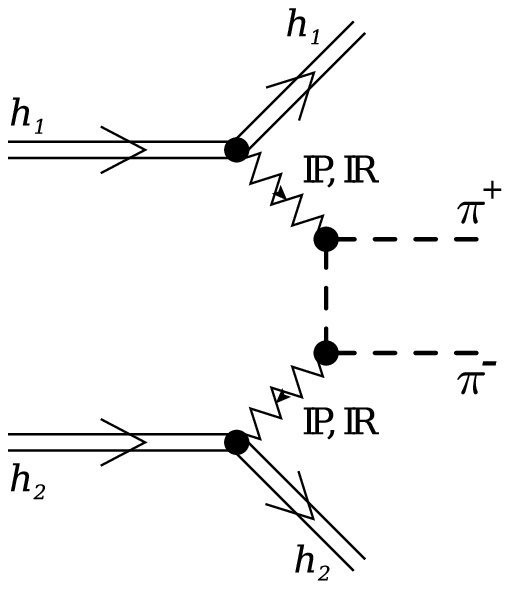}
  \includegraphics[width=4.5cm]{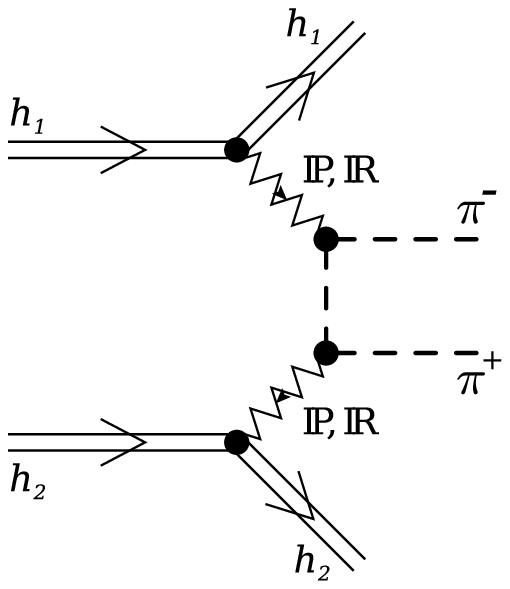}
\vspace{-0.7cm}
\end{center}
  \caption{A sketch of the dominant mechanisms of
exclusive production of the $\pi^+ \pi^-$ pairs
at high energies.
\label{fig:2pi_mechanisms}
}
\end{figure}

The underlying mechanism was proposed long ago in 
Ref.\cite{PH76}. The general situation is sketched 
in Fig.\ref{fig:2pi_mechanisms}.
The corresponding amplitude for the
$p p \to p p \pi^+ \pi^-$ process (with four-momenta 
$p_a + p_b \to p_1 + p_2 + p_3 + p_4$) can be written
as
\begin{eqnarray}
&&{\cal M}^{p p \to p p \pi \pi} \nonumber \\ 
&&= M_{13}(s_{13},t_1) \; F(t_a) \;  
\frac{1}{t_a - m_{\pi}^2} \;
F(t_a) \; M_{24}(s_{24},t_2) \nonumber \\
&&+ M_{14}(s_{14},t_1) \; F(t_b) \;
\frac{1}{t_b - m_{\pi}^2} \
F(t_b) \; M_{23}(s_{13},t_2)
 \;, \nonumber
\label{Regge_amplitude}
\end{eqnarray}
where $M_{ik}$ denotes "interaction" between nucleon $i$=1 
(forward nucleon) or $i$=2 (backward nucleon)
and one of the two pions $k=\pi^+$ (3), $\pi^-$ (4). 
In the Regge phenomenology they can be written as
a sums of two components:
\begin{equation}
M_{ik}(s_{ik},t_{1/2}) = 
M_{ik}^R(s_{ik},t_{1/2}) + M_{ik}^P(s_{ik},t_{1/2}) \; .
\end{equation}
The first terms describe the subleading reggeon exchanges 
while the second terms describe exchange of the 
leading (pomeron) trajectory.
The strength parameters of the $\pi N$ interaction
are taken from Ref.\cite{DL92}.
We choose post representation of the phenomenological 
exchange interaction, i.e. interaction for energy in the 
corresponding final state subsystem.
Above $s_{ik} = W_{ik}^2$, where $W_{ik}$ is the 
center-of-mass energy in the (i,k) subsystem.
More details of the calculation will be presented 
elsewhere \cite{LS09}.
The $2 \to 4$ amplitude (\ref{Regge_amplitude}) is used
to calculate the corresponding cross section including
limitations of the four-body phase-space.


\begin{figure}[!h]    %
\begin{center}
\includegraphics[width=6cm]{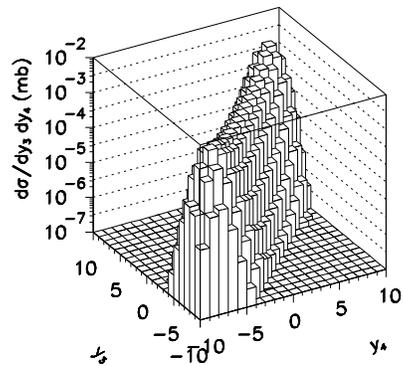}
\end{center}
\vspace{-0.7cm}
   \caption{\label{fig:map_y3y4.eps}
   \small 
Rapidity distribution of $\pi^+$ versus $\pi^-$
for W = 14 TeV.
}
\end{figure}

Here I show only one example of the 
two-dimensional distribution in rapidity of positively
charged pion and rapidity of negatively charged pion
at the LHC energy of $W =$ 14 TeV.
The distribution differs considerably from
the uniform population of the phase space.
One can see a two-dimensional shape of the ridge form
elongated along the line $y_3 = y_4$. The minimum of 
the cross section on the top of the ridge occurs when 
$y_3 = y_4 = 0$ and two maxima close to the phase space
ends. The minimum occurs in the part of the phase space
where the pomeron-pomeron contribution dominates, i.e. 
when both $W_{ik}$ are comparable and large.
The maxima are related to the dominance of
the pomeron-reggeon and reggeon-pomeron mechanisms, i.e.
where one of $W_{ik}$ is small and the second one is large.
The reggeon-reggeon contribution is completely negligible
which is due to the fact that both $W_{ik}$ cannot be
small simultaneously.
The ALICE collaboration at the LHC should be
able to measure such distributions.

\subsection{Exclusive $A A \to A A \rho^0 \rho^0$  \\
}

Exclusive production of elementary particles (lepton pairs,
Higgs, etc.) or mesons (vector mesons, pair of pseudoscalar
mesons, etc.) in ultrarelativistic heavy-ion collisions is 
an interesting 
and quickly growing field \cite{BGMS75,BHTSK02,Hencken} of 
theoretical investigation.
On experimental side the situation is slightly different.
So far only single-$\rho^0$ exclusive cross section
$A A \to A A \rho^0$ was measured \cite{STAR_rho0}. 


\begin{figure}[!h]   
\begin{center}
\includegraphics[width=4cm]{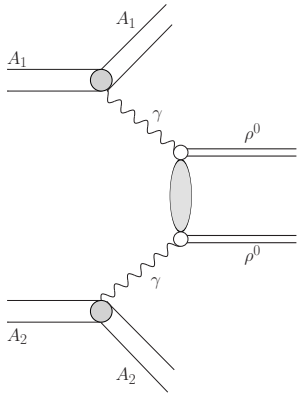}
\end{center}
\vspace{-0.7cm}
   \caption{\label{fig:diagram}
   \small 
The reaction discussed in this paper.
}
\end{figure}

Let us consider the process $A A \to A A \rho^0 \rho^0$
depicted in Fig.\ref{fig:diagram}.
The cross section takes the familiar form of a convolution
of equivalent photon fluxes and 
$\gamma \gamma$--cross sections.
Often flux factors of equivalent, almost on-shell, 
photons are calculated as for point-like particles 
with rescaled charge $e \to Z e$,
and the total cross section is calulated using
a simple parton--model type formula:
\begin{eqnarray}
\sigma \left( AA \to A (\rho^0 \rho^0) A \right) = \nonumber\\
\int d \omega_1 d \omega_2 
\frac{n(\omega_1)}{\omega_1}
\frac{n(\omega_2)}{\omega_2} 
\hat{\sigma} \left( \gamma \gamma \to \rho^0 \rho^0 \right) \; .
\label{EPA_formula}
\end{eqnarray}
The formulae (\ref{EPA_formula})
clearly does not take 
into account absorption effects when initial nuclei 
undergo nuclear breakup. This can be done in 
the impact parameter space where the geometry of the
collision is explicit.
Then rather two-dimensional flux factors \cite{Jackson} 
must be used.

The simple EPA formula can be generalized to
\begin{eqnarray}
&&\sigma \left( AA \to A (\rho^0 \rho^0) A \right) = \\
&&\int d^2 b_1 d \omega_1 d^2 b_2 d \omega_2 
N(\omega_1, b_1)
N(\omega_2, b_2) \; 
\nonumber \\
&& \theta 
\left(|\vec{b}_1 - \vec{b}_2| - R_{12} \right) \;
{\hat \sigma} 
\left( \gamma \gamma \to \rho^0 \rho^0 \right) \; .
\label{bspace_EPA_formula}
\end{eqnarray}
Here the extra $\theta$ function excludes those cases when 
nuclear collisions, leading to nuclear disintegration, 
take place ($R_{12} = R_1 + R_2$).
The two-dimensional fluxes in (\ref{bspace_EPA_formula}) 
are calculated in terms of the charge form factor
of nucleus \cite{BF91} as:
\begin{equation}
N(w,b) = \frac{Z^2 \alpha}{\pi^2} \Phi(x,b) \; ,
\label{2dim_flux}
\end{equation}
where the auxiliary function $\Phi$ reads: 
\begin{equation}
\Phi(x,b) = \Big|
\int_{0}^{\infty} du \; u^2 J_1(u) \;
\frac{F(-(x^2+u^2)/b^2)}{x^2 + u^2} 
\Big|^2  \; .
\label{auxiliary_Phi}
\end{equation}
The second ingredient of our approach is 
the $\gamma \gamma \to \rho^0 \rho^0$ cross section.
Here the situation is not well established.
The cross section for this process was measured
up to $W_{\gamma \gamma}$ = 4 GeV \cite{MPW}.
At low energy one observes a huge increase of the cross section.

\begin{figure}[!h]   
\begin{center}
\includegraphics[width=0.4\textwidth]{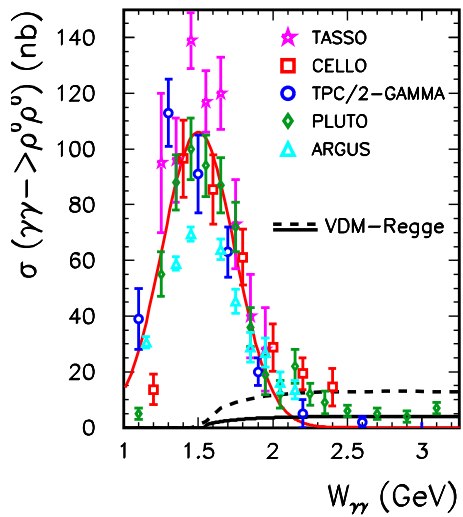}
\end{center}
\vspace{-0.7cm}
   \caption{\label{fig:fit_gamgam_rhorho}
   \small
The elementary cross section for 
the $\gamma \gamma \to \rho^0 \rho^0$ reaction.
We display the collection of the $e^+ e^-$ 
experimental data \cite{MPW} and our fit.
We show also our predictions based on the VDM-Regge
model decribed in \cite{KSS09}. For comparison we show also result when
the form factor correcting for off-shell effect is ignored
(see \cite{KSS09}).
}
\end{figure}

In Fig.\ref{fig:fit_gamgam_rhorho} we have collected 
the world data (see \cite{MPW} 
and references therein).
We use rather directly experimental data in order 
to evaluate the cross section in nucleus-nucleus 
collisions. Fig.\ref{fig:fit_gamgam_rhorho} 
shows our fit to the world data.


\begin{figure}[!h]  
\begin{center}
\includegraphics[width=0.4\textwidth]{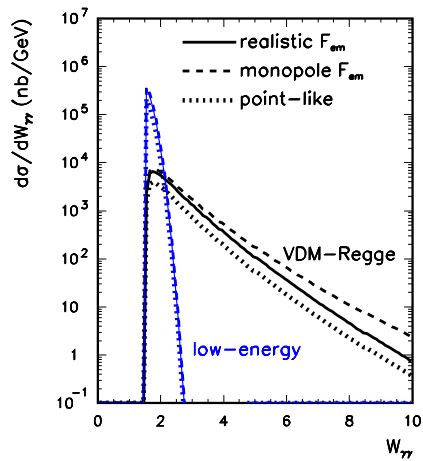}
\end{center}
\vspace{-0.7cm}
   \caption{\label{fig:dsig_dW} 
  \small 
The $Au + Au \to Au + Au + \rho^0 \rho^0$ cross section
as a function of $W_{\gamma \gamma}$ = $M_{\rho \rho}$
for the RHIC energy $\sqrt{s}_{NN}$ = 200 GeV.
The low- and high-energy components are shown separately.
}
\end{figure}

The cross section above $W$ = 4 GeV was never measured
in the past. It is well known that the cross section 
for $\gamma \gamma \to$ hadrons can be well described
in the VDM-Regge type model.
We use a similar approach for the
final state channel $\rho^0 \rho^0$.
In Fig.\ref{fig:fit_gamgam_rhorho} I present
the corresponding $t$-integrated cross section together
with existing experimental data taken from 
\cite{MPW}.
The vanishing of the VDM-Regge cross section at 
$W_{\gamma \gamma} = 2 m_{\rho}$ is due to 
$t_{min}$, $t_{max}$ limitations.
It is obvious from Fig.\ref{fig:fit_gamgam_rhorho} that the VDM-Regge 
model cannot explain the huge close-to-threshold enhancement. 
In Fig.\ref{fig:dsig_dW} we show distribution
of the cross section for the nucleus-nucleus scattering
in photon-photon center-of-mass energy for
both low-energy component and high-energy 
VDM-Regge component. Below $W$ = 2 GeV the low-energy
component dominates. The situation reverses above
$W$ = 2 GeV. One could study the high-energy component
by imposing an extra cut on $M_{\rho \rho}$.
However, the cross section drops quickly with increasing
invariant mass of the $\rho \rho$ pair.
The point-like in this (and following) figure(s) means the calculation
which excludes regions: $b_1 < R_{Au}$ and $b_2 < R_{Au}$.


\begin{figure}[!h]   
\begin{center}
\includegraphics[width=0.4\textwidth]{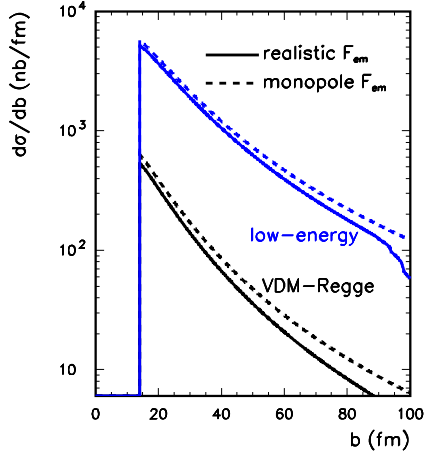}
\end{center}
\vspace{-0.7cm}
   \caption{\label{fig:dsig_dbm}
   \small 
The $Au + Au \to Au + Au + \rho^0 \rho^0$ cross section
as a function of the impact parameter $b$ for 
$\sqrt{s}_{NN}$ = 200 GeV. The meaning of the curves is
the same as in Fig.\ref{fig:dsig_dW}.
The cut off for $R_{12} \approx$ 14 fm is clearly visible.
}
\end{figure}

For illustration in Fig.\ref{fig:dsig_dbm} I show 
the model distribution in impact parameter 
$b = |\vec{b}_1 - \vec{b}_2|$.
Both distributions for the low- and high-energy
components are shown separately. I also show distributions
for "point-like" charge, monopole form factor
and realistic charge density (see \cite{KSS09}).
One can see slightly different results for different 
approaches how to calculate flux factors of equivalent
photons.


\begin{figure}[!h]  
\includegraphics[width=0.4\textwidth]{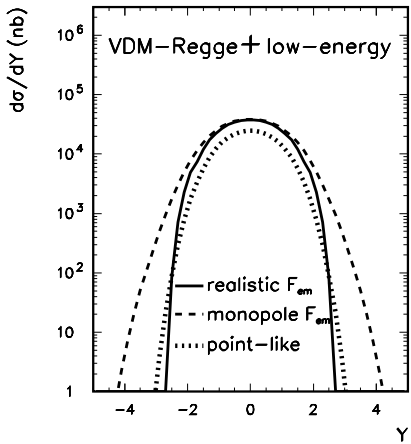}
\vspace{-0.7cm}
   \caption{\label{fig:dsig_dY}
   \small 
The $Au + Au \to Au + Au + \rho^0 \rho^0$ cross section
as a function of the rapidity of the $\rho^0 \rho^0$ pair 
$Y$ for $\sqrt{s}_{NN}$ = 200 GeV.
The meaning of the curves is
the same as in Fig.\ref{fig:dsig_dW}
}
\end{figure}

Finally in Fig.\ref{fig:dsig_dY} I show distribution
in rapidity of the $\rho^0 \rho^0$ pair.
Compared to the monopole form factor (usually used in the literature), 
the distribution obtained with realistic charge density is concentrated
at midrapidities, and configurations when both $\rho^0$'s
are in very forward or both $\rho^0$'s are in very backward
directions are strongly damped. A similar effect can be expected
for the $AA \to AA \mu^+ \mu^-$ reaction and could be studied
by the CMS and ALICE collaborations.


\vspace{0.3cm}

{\bf Acknowledgments}
The results presented here were obtained in collaboration with 
Wolfgang Sch\"afer, Roman Pasechnik, Oleg Teryaev, 
Mariola K{\l}usek and Piotr Lebiedowicz. 
This work was partially supported by the 
Polish Ministry of Science and Higher Education
under grants no. N N202 249235 and N N202 078735.

\end{document}